\documentclass[reprint,prb,superscriptaddress,showpacs]{revtex4-2}%
\usepackage{natbib} 

\setcitestyle{maxcitenames=6}
\usepackage{graphicx}
\usepackage{color}
\usepackage{amsmath}
\usepackage{amsfonts}
\usepackage{amssymb}
\usepackage{framed}
\usepackage{float}
\usepackage{placeins}
\usepackage[normalem]{ulem}
\usepackage{lineno}
\usepackage{lipsum}
\providecommand{\U}[1]{\protect\rule{.1in}{.1in}}
\makeatletter
\renewcommand*{\fnum@figure}{{\normalfont\bfseries \figurename~\thefigure}}
\renewcommand*{\@caption@fignum@sep}{\normalfont\textbf{ : }}
\makeatother
\usepackage{hyperref}
\hypersetup{
    colorlinks=true,
    linkcolor=blue,
    filecolor=magenta,      
    urlcolor=cyan,
}
\begin{document}
\title{Origin of spin reorientation and intrinsic anomalous Hall effect in the kagome ferrimagnet TbMn$_6$Sn$_6$}
\author{D. Connor Jones}
\thanks{equal contribution}
\affiliation{Department of Physics and Astronomy, George Mason University, Fairfax, Virginia 22030, USA}
\affiliation{Quantum Science and Engineering Center, George Mason University, Fairfax, Virginia 22030, USA}
\author{Suvadip Das}
\thanks{equal contribution}
\affiliation{Department of Physics and Astronomy, George Mason University, Fairfax, Virginia 22030, USA}
\affiliation{Quantum Science and Engineering Center, George Mason University, Fairfax, Virginia 22030, USA}
\affiliation{Department of Physics, Birla Institute of Technology and Science Pilani Hyderabad Campus, Hyderabad, Telangana 500078, India}
\author{Hari Bhandari}
\thanks{equal contribution}
\affiliation{Department of Physics and Astronomy, George Mason University, Fairfax, Virginia 22030, USA}
\affiliation{Quantum Science and Engineering Center, George Mason University, Fairfax, Virginia 22030, USA}
\author{Xiaoxiong Liu}
\affiliation{Department of Physics, University of Zurich, Winterthurerstrasse 190, Zurich CH-8057, Switzerland}
\author{Peter Siegfried}
\affiliation{Department of Physics and Astronomy, George Mason University, Fairfax, Virginia 22030, USA}
\affiliation{Quantum Science and Engineering Center, George Mason University, Fairfax, Virginia 22030, USA}
\author{Madhav P. Ghimire}
\affiliation{Central Department of Physics, Tribhuvan University, Kirtipur, Kathmandu 44613,, Nepal}
\affiliation{Leibniz Institute for Solid State and Materials Research, IFW Dresden, Helmholtzstr. 20, D01069 Dresden, Germany}
\author{Stepan S. Tsirkin}
\affiliation{Department of Physics, University of Zurich, Winterthurerstrasse 190, Zurich CH-8057, Switzerland}
\affiliation{Centro de F{\'i}sica de Materiales, Universidad del Pa{\'i}s Vasco, 20018 San Sebasti{\'a}n, Spain}
\affiliation{Ikerbasque Foundation, 48013 Bilbao, Spain}
\author{I. I. Mazin}
\affiliation{Department of Physics and Astronomy, George Mason University, Fairfax, Virginia 22030, USA}
\affiliation{Quantum Science and Engineering Center, George Mason University, Fairfax, Virginia 22030, USA}
\author{Nirmal J. Ghimire}
\thanks{corresponding author}
\email{nghimire@nd.edu}
\affiliation{Department of Physics and Astronomy, George Mason University, Fairfax, Virginia 22030, USA}
\affiliation{Quantum Science and Engineering Center, George Mason University, Fairfax, Virginia 22030, USA}
\affiliation{Department of Physics and Astronomy, University of Notre Dame, Notre Dame, Indiana 46556, USA}
\affiliation{Stavropoulos Center for Complex Quantum Matter, University of Notre Dame, Notre Dame, Indiana 46556, USA}
\begin{abstract}
TbMn$_6$Sn$_6$ has attracted a lot of recent interest for a variety of 
reasons, most importantly, because of the hypothesis that it may support
quantum-limit Chern topological magnetism, derived from the kagome geometry.
Besides, TbMn$_6$Sn$_6$ features a highly unusual magnetic reorientation 
transition about 100 K below the Curie point, whereby all spins in the system,
remaining collinear, rotate by 90$^\circ$. In this work, we address both issues 
combining experiment, mean-field theory and first-principle calculations.
Both magnetic reorientation and the unusual temperature dependence of the 
anomalous Hall conductivity (AHC)
find quantitative explanation in the fact that Mn and Tb, by virtue of the Mermin-Wagner theorem, have very different spin dynamics, 
with Tb spins experiencing much more rapid fluctuation. We were able to cleanly
extract the intrinsic AHC from our experiment, and calculated the same microscopically, with good semiquantitative agreement. We have identified 
the points in the band structure responsible for the AHC and showed that they
are not the kagome-derived Dirac points at the K-corner of the Brillouin zone,
as conjectured previously.
 \end{abstract}

\maketitle

\section*{Introduction}
The kagome lattice, a two-dimensional network composed of corner-sharing triangles, can harbor complex magnetic and electronic properties, including frustrated,
 non-collinear and non-coplanar spins \cite{Balents2010,Chen2014,Sachdev1992b,Hirschberger2019}, interesting electronic features, such as flat bands and 
Dirac 
points (DP), quantum spin liquids, and integer and fractional quantum Hall states \cite{Mazin2014c,Bolens2019,Guo2009b,Zhang2011e,Xu2015k,Tang2011}. 

\begin{figure*}[!ht]
\begin{center}
\includegraphics[width=.7\linewidth]{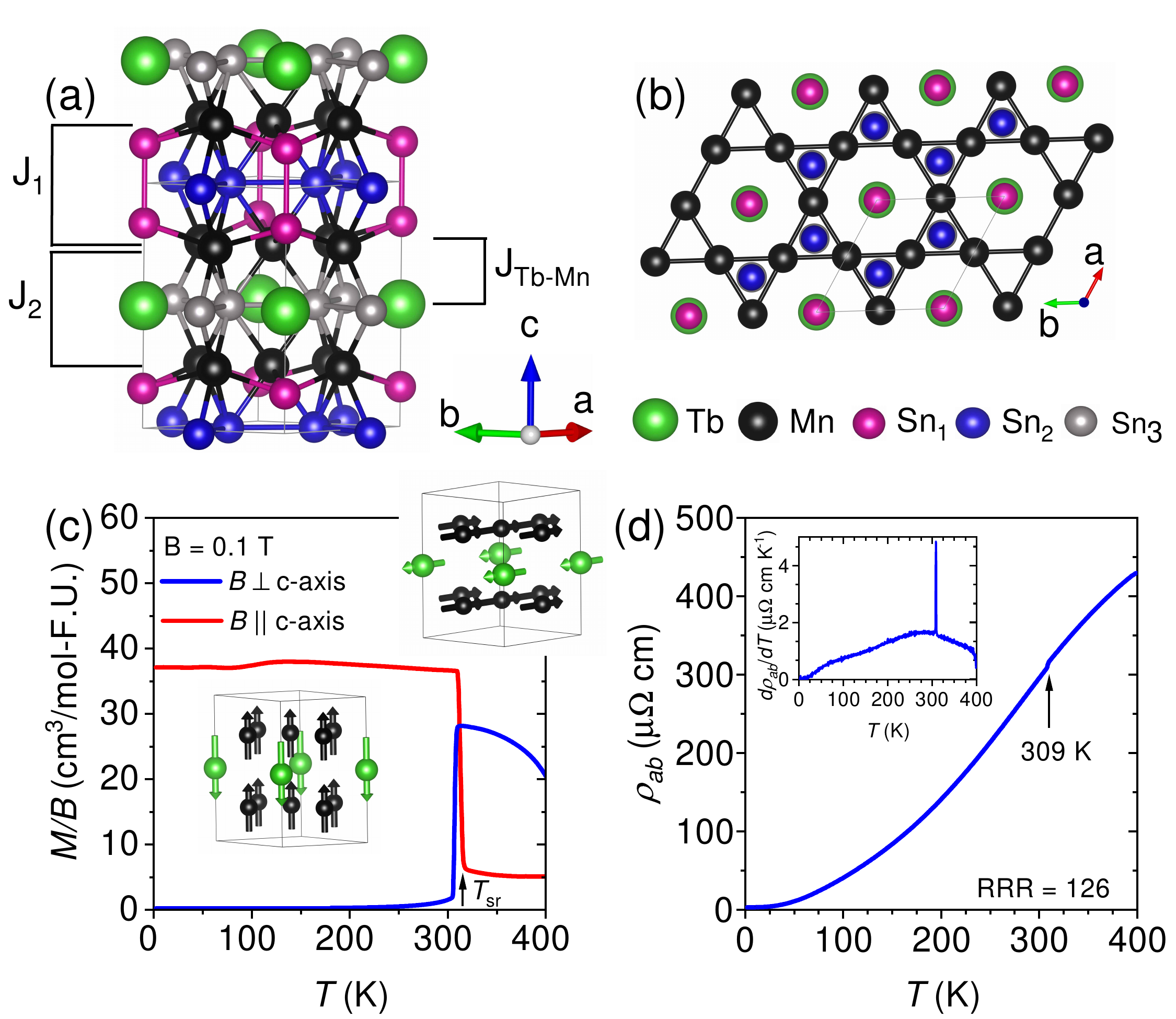}
\end{center}
\caption{Crystal structure and characterization of 
 TbMn$_{6}$Sn$_{6}$. (a) Sketch of crystal structure of TbMn$_{6}$Sn$_{6}$. The symbols $J_i$ are exchange constants between different Mn layers. $J_{Tb-Mn}$ is the exchange constant between Tb and Mn layers. (b) Top view of the structure shown in (a) within a unit cell shown by the grey solid lines where there are two kagome planes of Mn atoms with the formula Mn$_3$Sn separated by Sn$_3$ and TbSn$_2$ layers. (c). Field-cooled (FC) magnetic susceptibility of TbMn$_{6}$Sn$_{6}$ as a function of temperature measured parallel (red curve) and perpendicular (blue curve) to the crystallographic $c$-axis. At $T_{sr}$ = 309 K, the susceptibilities abruptly change directions indicating reorientation of the spins as illustrated by the two cubic graphics in the inset. The orientation of the Tb (green)  and Mn (black) magnetic moments above and below $T_{sr}$ from right to left, respectively. (d) Electrical resistivity of TbMn$_{6}$Sn$_{6}$ as a function of temperature. Inset shows the temperature derivative of the resistivity revealing the spin-reorientation transition at 309 K in the resistivity measurement.}%
\label{F1}%
\end{figure*}

Recently there has been growing interest in the three-dimensional (3D) compounds consisting of kagome nets of magnetic atoms. Their
symmetry-protected Dirac crossings can, by virtue of the exchange splitting, become Weyl points, and after inclusion of spin-orbit coupling 
may give rise to interesting topological properties \cite{Mazin2014c,Bolens2019,Guo2009b,Zhang2011e,Xu2015k,Tang2011}. $R$Mn$_6$Sn$_6$ 
($R166$), where  $R$ is a rare-earth element, are a recent addition to 
this class of compounds \cite{Asaba2020,Yin2020,Ghimire2021,Ma2021,Dhakal2021,Venturini2021,Zeng2022}. There, the Mn atoms form the kagome net in the 
basal plane of the hexagonal crystal structure as shown in Fig. \ref{F1}(a,b). The $R$166 compounds are rich in 
magnetic phases unlike other kagome magnets, due to their unique crystal 
structure \cite{Clatterbuck1999,Rosenfeld2008a,Ghimire2021,Siegfried2021,Dally2021}, making these compounds an excellent platform
to investigate the interplay of magnetism and the electronic structure. 

While the well-known 2D Kagome cuprates \cite{Shaginyan2020,science.aay0668} (herbertsmithite $etc$.) have been attracting a lot of interest due to their in-plane
magnetic frustration and proximity to spin liquids, 3D stannates are dramatically different in their physical properties. Those cuprates are both electronically
and magnetically extremely 2D. Their magnetic interactions are dominated by the frustrated antiferromagnetic (AF) in-plane Heisenberg exchange, while 
the electronic structure is well mappable onto a single-orbital nearest-neighbor tight binding model \cite{Mazin2014c}. On the contrary,
 the stannates are good metals
where the in-plane interactions are driven by kinetic exchange and therefore strongly ferromagnetic (FM) (no in-plane frustration),  and their electronic
structure is multiorbital and very 3D like. As a result, their interplanar magnetic interactions are long-ranged and of variable sign, and their electronic 
structure is complex and dispersive in all three directions, with only mild reminiscences of the notorious Dirac cones and flat bands, characteristic
of the single-orbital 2D Kagome model (the latter are known to be particularly fragile and rather far from true ``flatteness'' even in herbertsmithite-type compounds \cite{Mazin2014c}). This makes the physics and topology of stannates, and especially this 166 family, much more complex than that of this ``zero'' model, but also much more interesting. Their magnetic phase diagrams are extremely rich, and, with an exception of YMn$_6$Sn$_6$ (Y166), poorly understood. 
Some of them show unconventional
Hall behavior \cite{Wang2021,Ghimire2021}, including fluctuation-driven topological Hall effect (in Y166) \cite{Ghimire2021} and large anomalous
 Hall conductivity \cite{Asaba2020,Yin2020,Dhakal2021,Zeng2022}. 

{TbMn$_6$Sn$_6$ (Tb166) orders in a collinear in-plane ferrimagnetic (FiM) spin structure below 423 K \cite{Idrissi1991,Clatterbuck1999}. Highly unusual, with the
 decrease in temperature, at $T_{sr}$ = 309 K it experiences a sudden spin reorientation, where the moments spontaneously 
rotate from the basal plane toward the $c$-axis, and quickly set entirely along $c$, as can be seen in the magnetic susceptibility anisotropy, and illustrated by
 the sketch of the spin structure in Fig. \ref{F1}(c). All Mn spins are parallel to each other and antiparallel to the Tb spins. Tb is tri-valent, $i.e.,$ isovalent with Y,
 and has 8 f-electrons with the total magnetic moment of 9 $\mu_B$ (6 from the spins and 3 from the orbital moments). Additionally, Tb166 is a good metal with
 a large residual resistivity ratio (RRR), which is 126 in our sample as shown in Fig. \ref{F1}(d).} 
 
{Recently, tunneling experiments \cite{Yin2020} revealed a surface band of Tb166 with a gapped DP (``Chern gap'') at about 130 meV above the Fermi energy ($E_F$), which in a pure 2D case would have provided for an anomalous Hall conductivity (AHC) comparable 
with the experiment. Based on their band structure calculations, Ref. \cite{Yin2020} suggested that it was located at the K point in the Brillouin zone.
However, the calculated bands presented there had an implicit offset of the Fermi energy ($E_F$) by about 0.5 eV (see 
a detailed discussion in Ref. \cite{Ke}), which can hardly be the case in the bulk.
On the other hand, anomalous Hall effect is a bulk property that comes
 from the Berry curvature of all the occupied states see Eq.~\ref{eq2} in the Methods Section). Its estimation requires full calculations of the Berry curvature, taking into account $all$ electronic states, 
and not only those in the vicinity of one particular point in the Brillouin zone (BZ). It thus calls for a detail analysis of the band structure and calculations of the
 contributions of these bands to the AHC. Also, a detailed understanding of the role of the element Tb in establishing the collinear FiM order and spin reorientation transition is
 essential. A comparison of both magnetic and magnetotransport properties with the sister compound Y166 is extremely helpful.} 

{In this paper, we investigate these three issues by combining experimental observations and first principles calculations. First, we show that different dynamics of spin fluctuations 
 on Tb and on Mn plays a key role in the spin reorientation process. We then show that although Tb166 exhibits a large intrinsic AHC, the separation of its intrinsic and
 extrinsic contributions is quite complex and in this scenario, spin dynamics is crucially important and should be taken into account in analyzing the AHC in
 $R$166 compounds. Our calculations reveal that the AHC in Tb166 results from a combination of multiple factors and states across different regions of 
 the 3D BZ, rather than from one particular point in the BZ. We find that contrary to the ``zero model'', the near-Fermi level DP in Tb166, while can still be, with some imagination, identified at the K-H line in the BZ, are highly dispersive; the band that generates the largest contribution to the AHC in our calculations 
 is derived from Mn d$_{z^2-1}$ orbitals and is located around 1 eV above the Fermi level at K point and around 1 eV 
below it at H point. The two DPs closest to the Fermi level at the K-point (see Ref. \cite{Ke} for a comprehensive discussion of high-symmetry DPs), at 
$\approx -50$ and $\approx +230$  meV, contribute basically nothing, nor does the DP identified in \cite{Yin2020}, which is situated 
at $\approx +700$ meV in the bulk. A significant contribution to the AHC comes from where the Dirac lines cross the Fermi level (one for spin up and the other for spin-down), which 
occurs at $k_z\approx 0.25$ r.l.u. (reciprocal lattice units).
Additionally, there are several accidental (some of them tilted) DPs near the  
Fermi level, which also contribute to the Berry curvature and Hall conductivity. Amongst them, the most notable is the spin-down 
Dirac line crossing the Fermi level at the midpoint between A and L ($k_z$=0.5 r.l.u.), with some contribution from an accidental tilted spin-up DP between
K and $\Gamma$.}

\section*{Methods}
Single crystals of TbMn$_6$Sn$_6$ were grown by a self-flux method using excess Sn as the flux. Tb pieces (Alfa Aesar 99.9 \%), Mn pieces (Alfa Aesar 99.95 \%), and Sn shots (Alfa Aesar 99.999 \%) were loaded to a 2 mL aluminum oxide crucible in a molar ratio of 1:6:20 and sealed in a fused silica ampule under vacuum. The sealed ampule with the crucible was heated to 1150 $^\circ$C for 10 hours, homogenized at 1150 $^\circ$C for 12 hours, and then cooled to 650 $^\circ$C at 4 $^\circ$C/hour. Once the furnace reached 650 $^\circ$C, the excess Sn-flux was decanted using a centrifuge. Many well faceted hexagonal single crystals were obtained in the crucible. The crystal structure was verified using powder x-ray diffraction at room temperature using a Rigaku MiniFlex diffractometer. A small amount of the crystals from each batch were ground into powder. These powder samples were used to collect x-ray diffraction patterns, (shown in Supplementary Fig. S1) for one representative batch, using Rietveld refinement \cite{Mccusker1999} with the FULLPROF software \cite{Rodriguez-carvajal1993}. The results of the refinement are presented in Supplementary Table S1.  Magnetic and transport measurements were performed on single crystals oriented for $c$-axis. 

DC magnetization, resistivity, and Hall measurements were performed in a Quantum Design Dynacool Physical Property Measurement System (PPMS) with a 9 T magnet. The AC Measurement System (ACMS) option was used for the DC magnetization measurements. The saturation magnetization and coercive fields were calculated by averaging between the heights and widths of the hysteresis curves, respectively, from the magnetization data with the magnetic field $B$ parallel to $c$ axis. The resistivity and Hall measurements employed the conventional four-probe method by attaching 25 $\mu$m diameter platinum wires with Epotek H20E silver epoxy. 2 mA of electrical current was used for the transport measurements.  The contact misalignment, in magnetoresistance and Hall  measurements, was corrected by  field symmetrization and antisymmetrization of the measured data, respectively.

The first principles calculations for electronic and magnetic structure of Tb166 were performed utilizing the local density approximation with the Perdew-Burke-Ernzerhof (PBE) gradient correction, (GGA) for the exchange-correlation functional, orthogonal plane-wave basis sets and the pseudopotential method to account for electron-ion interactions as implemented in the integrated suite for electronic structure calculation Quantum Espresso \cite{QE2017}. The Tb pseudopotential in our calculations include the 4f electrons within the open-core model. Note that inclusion of the 4f electrons in the valence configuration, with a sizeable Hubbard interaction U for f electrons, does not alter the low energy electronic structure significantly. The single-particle wave functions were evaluated using a plane-wave energy cutoff of 600 Ry. We utilized the Wannier90 \cite{Wannier1} program to generate maximally localized Wannier functions (MLWFs) from the Bloch states by convolution with unitary matrices and minimizing the spread of the MLWFs in real space. The method is independent of the choice of basis sets for the Bloch functions and the locality of MLWFs could be exploited to derive accurate band structures and low energy Fermi surface properties at considerably less computational cost. Note that we have included the effect of spin-orbit coupling in our first-principles calculations during the evaluation of the anomalous Hall conductivity at the level of constructing the maximally localized Wannier functions. Adopting the Kubo formalism, the intrinsic contribution to the anomalous Hall conductivity can be written as:~\cite{Xiao, Sinova}
\begin{align}
\sigma ^{AHE}_{\alpha \beta }=\frac{e^{^{2}}}{\hbar} \sum_{n\neq n^{'}} \int \frac{d^{3}\textbf{k}}{(2\pi)^{3}}[f(\varepsilon _{n}(\textbf{k}))-f(\varepsilon _{n^{'}}(\textbf{k}))]\nonumber \\
\times \; \textit{Im} \; \frac{\left< n,\textbf{k}\left|v_{\alpha}(\textbf{k}) \right|n^{'},\textbf{k}\right>\left< n^{'},\textbf{k}\left|v_{\beta}(\textbf{k}) \right|n,\textbf{k}\right>}{[\varepsilon _{n}(\textbf{k})-\varepsilon _{n^{'}}(\textbf{k})]^{2}}  \label{eq1}
\end{align}
where $f_{nk}$ are the Fermi Dirac distribution functions. Equivalently, Eq. \ref{eq1} can be rewritten as:~\cite{Vanderbilt1, Vanderbilt2, Niu}
\begin{align}
\sigma^{AHE}_{\sigma \beta } = -\frac{e^{2}}{\hbar}\epsilon _{\alpha \beta \gamma}\int \frac{d^{3}\textbf{k}}{(2\pi)^{3}}\: \Omega_{\gamma}(\textbf{k})\nonumber\\ = -\frac{e^{2}}{\hbar}\sum_{n}\epsilon _{\alpha \beta \gamma}\int \frac{d^{3}\textbf{k}}{(2\pi)^{3}}\: f_{n,\textbf{k}}\: \Omega_{\gamma}(n,\textbf{k}) \label{eq2}
\end{align}
Here, the anomalous Berry term is obtained by convolution of the Berry curvature with that of the Fermi function and summed over all the occupied states, where the Berry curvature can be written in terms of the Bloch functions as
\begin{align}
\Omega_{\gamma}(n,\textbf{k}) = i\left<\nabla_{\textbf{k}}u_{n}(\textbf{k})\left| \times \right| \nabla_{\textbf{k}}u_{n}(\textbf{k}) \right> 
\end{align}
While several electronic structure codes \cite{wanniertools, Wannier1} exist for evaluation of Berry curvature in real materials, we used the WannierBerri code \cite{Tsirkin2021} in order to reach high precision in the evaluation of AHC. Namely, our procedure incorporates minimal-distance replica selection method \cite{ Pizzi_2020} for accurate Wannier interpolation and the recursive adaptive refinement for accurate determination of physical quantities by the integration of rapidly oscillating functions in the k-space for evaluation of the anomalous Hall coefficient and appropriate consideration of symmetry properties. 
Selected calculations were verified against the full potential local-orbital (FPLO) code \cite{Koepernik}, where the linear tetrahedron method was employed with 12 $\times$ 12 $\times$ 6 k mesh.

\section*{Results and Discussion}

\subsection*{Magnetism}
We first discuss the ferrimagnetic (with all Mn spins parallel) rather than pseudo-antiferromagnetic state in Y166, and then the easy-axis magnetic 
anisotropy of Tb166 developed below the spin reorientation temperature ($T_{sr}=309$ K).
 We begin with the much better understood spiral ground state of Y166. It was appreciated decades ago that this spiral can be exceedingly well described in  
a three-parameters mean-field 1D Heisenberg chain model, where the Mn planes are ferromagnetically ordered, the interplanar exchange $via$ the Sn$_3$ layer, $J_1$ is ferromagnetic,
the one  $via$ the YSn$_2$ layer, $ J_2$, is antiferromagnetic, and  the second-neighbors interplanar exchange, $J_3$, is ferromagnetic \cite{Rosenfeld2008a} [Fig. \ref{F1}(a)]. We have demonstrated recently that this model, augmented with single-site anisotropy and anisotropic exchange, fully described the complex magnetic phase diagram 
in this compound\cite{Ghimire2021}. 

When Y is substituted with a magnetic ion, such as Tb, its magnetic interaction with the Mn plane must be taken into account. Regardless of its sign,
this interaction can be integrated out yielding an effective ferromagnetic interaction $\tilde J_2=-2J_{Mn-Tb}$ (here and below we absorb the $SS'$ factor into
the definition of $J$). This interaction
is rather strong and overwhelms the direct antiferromagnetic $J_2$ one. Given that $|J_3|\ll|J_2|,|J_1|$, the resulting magnetic Hamiltonian is
not frustrated, and the system readily orders ferrimagnetically 
(the actual sign of $J_{Mn-Tb}$ is antiferromagnetic \cite{Ke}). This is also fully 
consistent with our first principles calculations.

Next, we look at the magnetic anisotropy. It is instructive to compare the behavior of the 166 compounds across the entire series of rare earths. 
At the onset of magnetic ordering, all of them have easy plane anisotropy \cite{Clatterbuck1999}.  
[Except for the spiral-magnetic  $R$166 ($R$ = Sc, Y, Lu) and Tm166, they all form collinear ferrimagnetic structures]. The non-magnetic rare earths (Y, Lu 
and Sc) retain this easy plane anisotropy all the way down to low temperatures, and so does Gd, which has a full f-shell and no single-ion magnetic anisotropy. 
This clearly establishes the fact that Mn anisotropy 
(both single site and exchange) in 166 compounds is easy plane.  Er and Tm present an interesting case: they form antiferromagnetic (Er) or short-pitch spiral (Tm) structures, indicating that the transferred interaction  $\tilde J_2$ in these materials is weaker than $J_2$, so the net interaction is still antiferromagnetic. As a result, Mn$-R$ interaction is frustrated (and $R-R$ one very weak), so the $R$ sub-lattice remains disordered till rather low temperatures (75 and 58 K, respectively), and even after ordering does not couple with the Mn sub-lattice, and thus does not affect its anisotropy.

The most interesting situation emerges in Tb, Dy, Ho and Er. There, $\tilde J_2$ is strong enough to induce a collinear state. Given the hexagonal crystal field, the 
natural lowest-order magnetic anisotropy for these $R$ ions is uniaxial, and strong. Detailed calculations \cite{Ke} show that in all the cases the net 
 anisotropy {\it at zero temperature} is dominated by the $R$, and dictates the 
magnetization direction for the entire crystal.  The crystal field on the $R$ site being comparable with, or smaller than the SOC, the total energy 
is determined by the transformation of the corresponding spherical harmonics between the spin tilting angle and the crystallographic $z$ axis,
which for f-ions include terms of up to $M_z^6$ with comparable coefficients. Direct calculations \cite{Ke} have demonstrated that, indeed,
the magnetic anisotropy energies for (Tb--Er)166 include quartic, and likely sextic terms, which generate low-symmetry zero-temperature anisotropy in Ho166 and Dy166.

\begin{figure*}[!ht]
\begin{center}
\includegraphics[width=.7\linewidth]{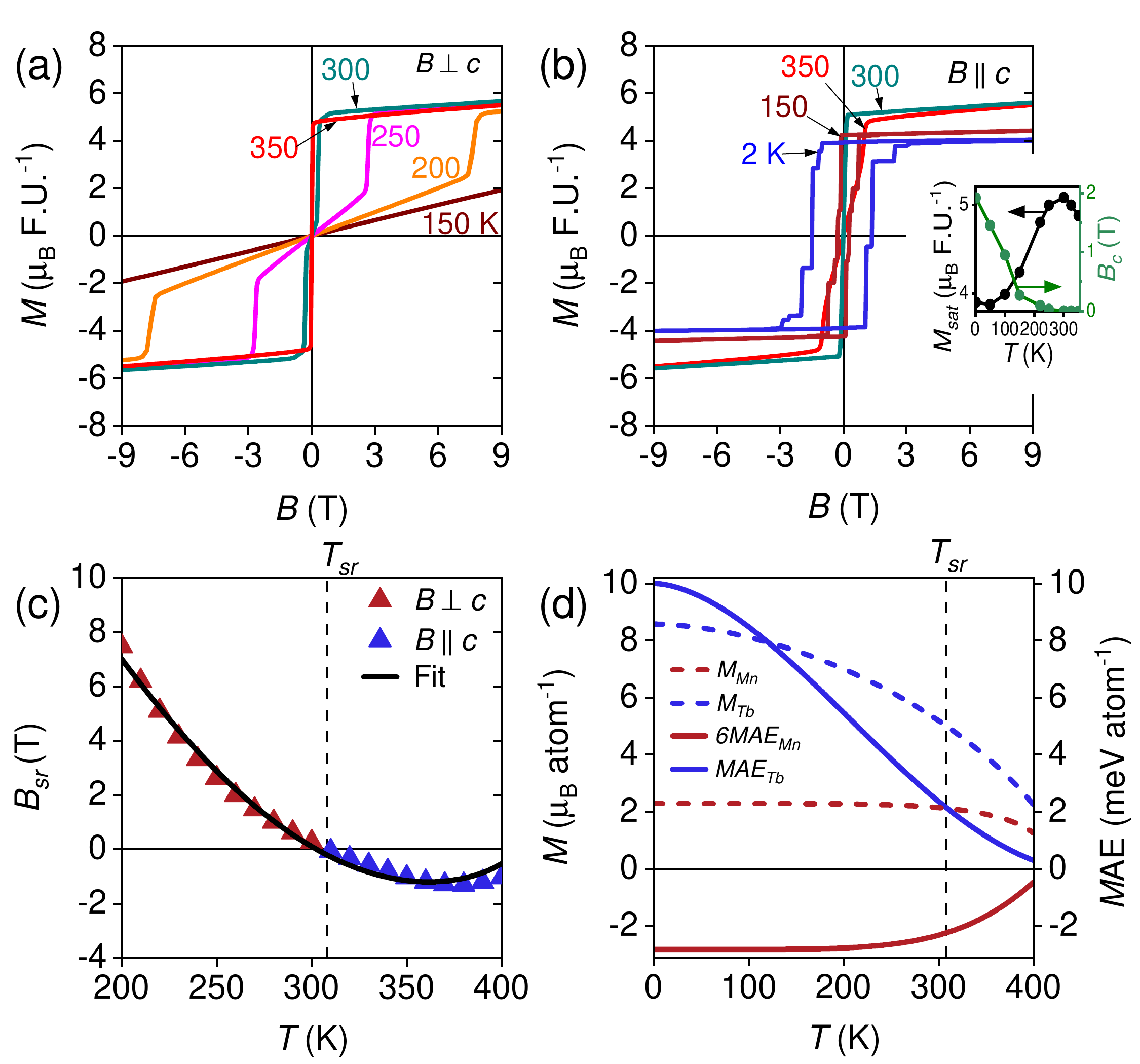}
\end{center}
\caption{ Magnetic field dependence of magnetization with a) $B \bot c$ and  b) $B \| c$. The inset in panel (b) shows the temperature dependence of saturation magnetization and the critical field. (c) Magnetic phase diagram for the spin-reorientation transition. Positive values for $B$ corresponds to $B \bot c$ whereas negative values correspond to $B \| c$. The fitted curve with $K_{Mn} = -8.96\times$10$^{-2}$ meV$\mu_B^{-2}$ and $K_{Tb}$ = 0.136 meV$\mu_B^{-2}$ is observed on top of the experimental values. The dashed vertical line represents the zero-field spin reorientation transition temperature $T_{sr}$. (d) Temperature dependence of the magnetic moments on Mn (red) and Tb (blue) (dashed lines) from Ref. \cite{Idrissi1991}, and magnetic anisotropy energies ($\mathcal{ MAE}$) of the Mn (red) and Tb (blue) sublattices (solid lines) [$\mathcal{ MAE}_{Mn}(0) = -0.47$ meV Mn$^{-1}$ and $\mathcal{ MAE}_{Mn}$(0) = 10 meV Tb$^{-1}$].}%
\label{F2}%
\end{figure*}

Let us now quantify this physical picture in reference to, specifically, Tb166, and address the reorientation transition at higher temperature.
First, let us show our experimental data.
The external field dependence of magnetization $M(B$) with $B\bot c$ and $B \| c$  is shown in Figs. \ref{F2}(a) and \ref{F2}(b), respectively. For $B\bot c$,
 the shape of the curves indicates a soft ferromagnetic behavior above $T_{sr}$ via a sharp increase at low fields followed by saturation at higher fields. 
Below $T_{sr}$, a metamagnetic transition is observed that corresponds to the spins flopping from the $c$-axis to the basal plane. The metamagnetic
 transition field increases with the decrease in temperature and  surpasses 9 T below 150 K.  For $B\| c$, the curves also display soft
 ferromagnetic behavior above 250 K. Even above $T_{sr}$, the spins are flopped from the basal plane with relatively smaller magnetic field.  Below 250 K 
a hard ferromagnetic behavior is observed via the emergence of hysteresis loops containing asymmetric steps, which increase in width with decreasing
 temperature attaining a coercive field of 2 T at 2 K [inset of Fig. \ref{F2}(b)]. The saturation magnetization $M_{sat}$ reaches approximately 3.9 $\mu_B$ F.U.$^{-1}$ at 2 K and increases up until around 
300 K [inset of Fig. \ref{F2}(b)]. The metamagnetic transition for $T < T_{sr}$ with $B\bot c$  [Fig. \ref{F2}(a)] and $T > T_{sr}$  with $B \| c$  [Fig. \ref{F2}(b)]
 (tracked by d$M$/d$B$ in Fig. S2) represents the critical field required to induce a spin-reorientation transition $B_{sr}$($T$) at a particular temperature.
 A critical field -temperature phase diagram is shown in Fig. \ref{F2}(c). At higher temperatures, the magnetocrystalline anisotropy along the kagome planes dominates, resulting in the “easy-plane” type ordering, similar to Y166 in the ground state. Below $T_{sr}$ the uniaxial anisotropy of the Tb-sublattice dominates. In addition to this zero field spin-reorientation, at any temperature, the spin-reorientation transition in Tb166 can also be induced by applying an external field $B_{sr}$ along the hard magnetization direction, that takes place through a first order magnetization process (FOMP).

Now let us rationalize and quantify these observations. The effective Heisenberg Hamiltonian for the  $\tilde J_2$ exchange reads:
\begin{eqnarray}
\label{Eq1 }
H_2 = -2|J_{Mn-Tb}|\cos(\theta/2),
\end{eqnarray}     
while the effective magnetic anisotropy Hamiltonian (absorbing, as usual, the anisotropic exchange into the single site anisotropy) is (see Supplementary Note 1 for details)
\begin{eqnarray}
\label{Eq2 }
E_{anis} = 6\mathcal{MAE}_{Mn} (T) + \mathcal{MAE}_{Tb} (T),
\end{eqnarray}
with \begin{eqnarray}
\label{Eq3 }
\mathcal{MAE}_i(T) = K_iM^2_i(T)\frac{M^2_i(0)}{3M^2_i(0)-3M_i(0)M_i(T)+M^2_i(T)},\nonumber
\end{eqnarray}
where  $\mathcal{MAE}$ is the $T$-dependent magnetic anisotropy energy, $M_i$ and $K_i$ are the ordered ($i.e.$, averaged over thermal fluctuations)
magnetic moments and 2$^{nd}$ order anisotropy coefficients 
for the two atoms and sublattices ($i$ = Mn, Tb), respectively. In the mean-field approximation 
 $M_{Mn}(T)$ and $M_{Tb}(T)$ [Fig. \ref{F2}(d)], can be described by the Brillouin functions. Importantly, the molecular (Weiss) mean field on the Mn sites 
is large and is equal to $6J_{in-plane}\sim 3000$ K $\gg T_C$, while that on the Tb site is much smaller, $6J_{Mn-Tb}\sim 600$ K \cite{Ke},
 comparable with $T_C$.
As a result, the ordered magnetic moment on the Tb site shown by blue dashed line in Fig. \ref{F2}(d), as probed by neutrons, exhibit a much more gradual temperature decline (essentially mean field),
while that on the Mn site (red dashed line) changes slowly (according to its larger Curie-Weiss temperature, $T_{CW}$), until it rapidly drops near $T_C$, which is strongly reduced from $T_{CW}$ by 2D (Mermin-Wagner) fluctuations. 

\begin{figure}[ht]
\begin{center}
\includegraphics[width=.95\linewidth]{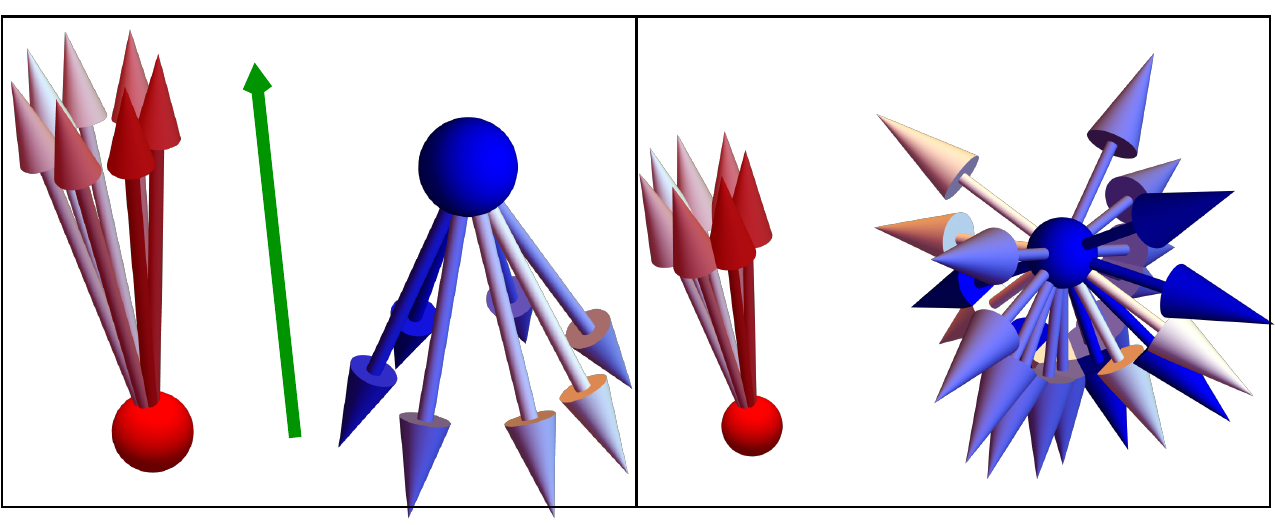}
\caption{Cartoon illustrating the spin dynamics of Mn (red) and Tb (blue) spins at low (left) and high (right) temperatures. The green arrow shows the net magnetization direction.}\label{cartoon}
\end{center}
\end{figure}

This is illustrated in Fig. \ref{cartoon}, showing fluctuating moments at $T\ll T_C$ and $T\alt T_C$. Correspondingly, the relative contribution of Tb into the overall anisotropy energy gets smaller as the temperature is increased and leads to the reorientation transition, above which the total anisotropy is dominated by Mn. Note that in this model, neglecting higher order anisotropy term, the transition is first order; this is not true in Dy166 and Ho166, where the higher order terms lead to an incomplete reorientation.

Fig. \ref{F2}(d) uses Curie-Weiss parameters fitted to existing  neutron powder diffraction data \cite{Idrissi1991}. As Fig. S3 shows, the fitting is very good.
By equating $E_{anis}$ [Eq. (2)] to the single-site Zeeman energy at $B$ = $B_{sr}$, given by
\begin{eqnarray}
\label{Eq4 }
E_{Zeeman}(T, B_{sr}) = [M_{Tb}(T) - 6M_{Mn}(T)]\mu_B\mu_0B_{sr}(T),\nonumber
\end{eqnarray}
where $\mu_B$ is the Bohr magneton, a fitted curve for $\mu_0B_{sr}(T)$ is generated in Fig. 2(c) with $K_{Mn} = - 8.96\times10^{-2}$ meV$\mu_B^{-2}$ and $K_{Tb}$ = 0.136 meV$\mu_B^{-2}$. The negative sign of $K_{Mn}$ indicates that the Mn sublattice favors the easy-plane magnetization direction. Conversely, the positive sign of $K_{Tb}$ indicates that the Tb sublattice favors the easy-magnetization direction along the $c$-axis due to a hexagonal crystal field splitting, as discussed above.
 Moreover, because Tb$^{3+}$ is a heavier ion, the magnitude of $K_{Tb}$ is significantly larger than $K_{Mn}$, thereby contributing more to $E_{anis}$ for a given $M_i$ \cite{Ke}.
 Therefore, the remarkable agreement of the fitted curve [Fig. \ref{F2}(c)] with the experimental data reveals that
 the spin-reorientation phase diagram is quantitatively described by the temperature dependencies of $M_{Mn}$ and $M_{Tb}$. The calculated values for $\mathcal{MAE}_{Mn}(T)$ and $\mathcal{MAE}_{Tb}(T)$
 are shown in Fig. \ref{F2}(d) by red and blue solid lines, respectively. Analogous to $M_{Mn}$ and $M_{Tb}$, respectively, 
$\mathcal{MAE}_{Mn}$ remains relatively constant from around 300 K down to zero temperature whereas $\mathcal{MAE}_{Tb}$ dramatically increases 
with decreasing temperature. At $T_{sr}$, $\mathcal{MAE}_{Mn}\sim -\mathcal{MAE}_{Tb}$, revealing that the spin reorientation transition at zero-field 
occurs when the two competing anisotropy energies of the two sublattice FiM system cancel out. In the ground state, $\mathcal{MAE}_{Mn} = - 0.47$
 meV per Mn and $\mathcal{MAE}_{Tb} = 10.0$ meV per Tb. To understand the contributions of Heisenberg exchange and single-site anisotropy terms within
 $\mathcal{MAE}_{Mn}$, we compare with the YMn$_6$Sn$_6$ compound. A recent study by Ghimire et al. \cite{Ghimire2021}
  estimated $\mathcal{MAE}_{Mn}$ for YMn$_6$Sn$_6$ in ground state to be $\sim -0.12$ meV per Mn atom. By assuming the single-site anisotropy on Mn- and Tb-sub-lattices coexist independently, we argue the larger magnitude of $\mathcal{MAE}_{Mn}$ for Tb166 arises from $J_{Mn-Tb}$ exchange. Recently, 
Lee et al. \cite{Ke} calculated magnetic anisotropy of Tb166 excluding Tb $f$-electrons to be $-1.7/2=-0.85$ meV. This large value can be assigned to the
anisotropy of the $d$ electrons on Tb, included in Ref. \cite{Ke} together with that of Mn. Please note that in Fig.  \ref{F2}(c) we only show spin-reorientation fields above $T\approx 200$ K. The reason is that the formalism presented above models single-site magnetic anisotropies as quadratic functions of the fluctuation-average moments. This is a good approximation for Mn d-electrons, but for Tb it is known\cite{Ke} to include large quartic and even sextic terms; this makes the quadratic approximation increasingly less reliable at low temperatures, where the average magnetic moment on Tb is large.

\subsection*{Anomalous Hall effect}
Now we discuss the anomalous Hall effect (AHE), which has attracted considerable attention recently in the $R$Mn$_6$Sn$_6$ compounds \cite{Asaba2020,Yin2020,Ghimire2021,Ma2021,Dhakal2021,Zeng2022}. An AHE is the transverse voltage induced by a longitudinal current flow
 in ferromagnetic materials without an external magnetic field. The AHE can have both intrinsic and extrinsic contributions. The former comes from the
 Berry curvature of the electronic bands, whereas the latter is related to the electron scattering effects such as side jumps and skew scattering. One of the
 biggest challenges in studying AHE is separating the intrinsic AHE from the extrinsic ones. In an ideal crystal at zero temperature the latter is zero, and so 
is the longitudinal resistivity $\rho_{xx}$ which led to the idea of using a protocol that 
relates  $\sigma^A_{xy} $ to  $\rho_{xx}$ $via$ a power expansion, 
\begin{eqnarray}
\label{Eq5}
\rho_{yx}^A &=& a + b\rho_{xx}+c\rho_{xx}^2+d\rho_{xx}^3+ ...\\ 
\label{Eq5a}
\sigma_{xy}^A &=& a\sigma_{xx}^{2} + b\sigma_{xx}+c+d/\sigma_{xx}+ ...,
\end{eqnarray}
with the idea that the free term $c$ in Eq. \ref{Eq5a} (or the coefficient of the quadratic term in Eq. \ref{Eq5}) will represent the intrinsic AHC (the terms higher than quadratic are neglected).
 It was rather soon realized, however, that the extrinsic AHC scales  differently
with  $\rho_{xx}$ of different origin. The widely discussed ones are skew scattering and side jumps. The contribution from the former is proportional to the resistivity, while that from the latter is proportional to the square of the resistivity, thus complicating the extraction of the intrinsic AHE.
Furthermore, if there 
are several mechanisms for residual resistivity (impurities,
defects, domain walls) the coefficients may be different for different mechanism (see a detailed discussion of this issue in Ref. \cite{Grigoryan}). In ferromagnetic materials 
scaling of the AHC can be influenced by phonon and magnon contributions that can vary significantly with temperature. While Ref. \cite{Crepieux2001} argued convincingly that  
phonon scattering does not contribute to the linear term in Eq. \ref{Eq5} (and equivalently in Eq. \ref{Eq5a}) because of the sign-changing potential fluctuations that phonons 
generate, it is not clear whether the 
same argument can be applied to transverse magnons, which represent fluctuations around a given average magnetic direction.
To the best of our knowledge, a theory of AHE in presence of fluctuating localized spins has never been developed.

Not surprisingly, given all these pitfalls and caveat,  in real life the simplified version of Eqs. \ref{Eq5} and \ref{Eq5a} that is ubiquitously used is: \cite{Tian2009,Ye2018c,Yin2020},
\begin{eqnarray}
\label{Eq6a}
\rho_{yx}^A(T)= a +c\rho_{xx}^2(T), \label{6a}\\ 
\label{Eq6b}
\mathrm{or,}~\sigma_{xy}^A(T)= a\sigma_{xx}^2+c,
\end{eqnarray}
Even a more sophisticated version  presented in Ref. \cite{Grigoryan} where $\rho_{yx}^A$ takes the form:
\begin{eqnarray}
\rho_{yx}^A(T)&=&\alpha_I\rho_{xx}^I\rho_{xx}(T)+\beta_I \rho_{xx}^I \nonumber\\
&+&\alpha_P\rho_{xx}^P\rho_{xx}(T)+\beta_P \rho_{xx}^P+\gamma\rho_{xx}^2(T), \label{G}
\end{eqnarray}
where everything but $\rho_{xx}(T)$ is a temperature-independent parameter characterizing different scattering mechanism ($\alpha_{I,P}$ are side-jump coefficients due to impurity/phonon scatterings, and $\beta_{I,P}$ are the skew-scattering coefficient due to impurity, and phonon-impurity cross scattering), often fails to describe the experimental data. 
\begin{figure*}[ht]
\centering
\includegraphics[width=1\linewidth]{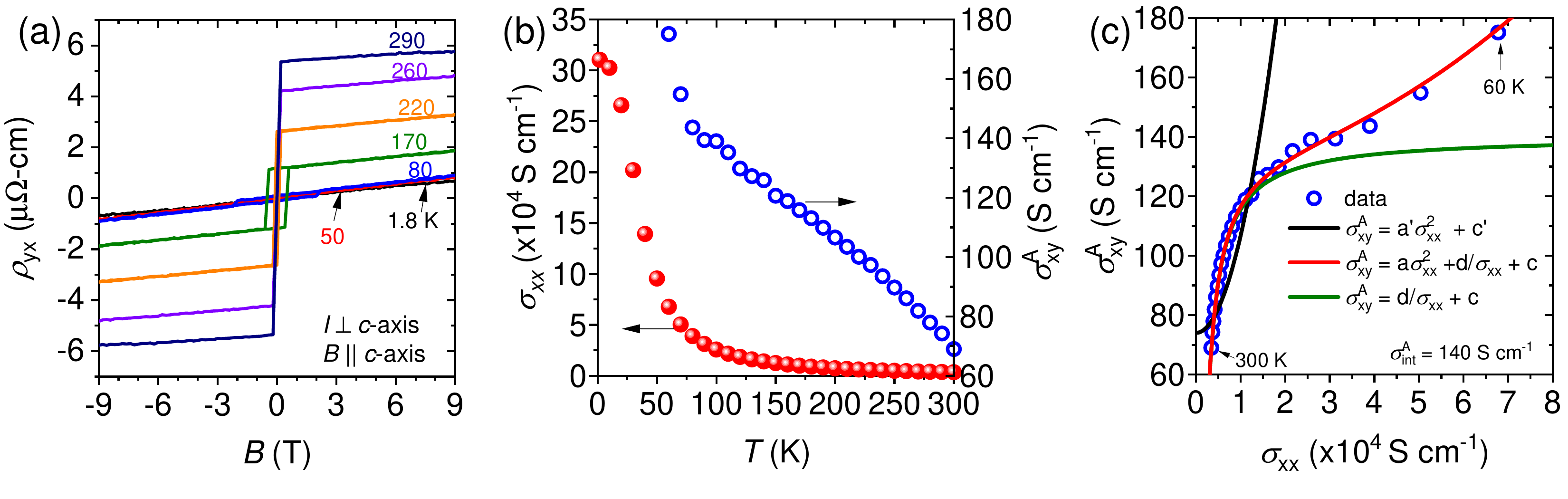}
\caption{a) Anomalous Hall resistivity ($\rho_{yx}$) of TbMn$_6$Sn$_6$ as a function of magnetic field ($B$) at some representative temperatures. b) Longitudinal conductivity ($\sigma_{xx}$, red spheres), and anomalous Hall conductivity ($\sigma_{xy}^A$, blue circles) as a function of temperature.  c) $\sigma_{xy}^A$ (blue circles) as a function of $\sigma_{xx}$ between 300 K (bottom left) and 60 K (top right). The solid lines are the plot of different functions shown in the legend. a, c and d's are constants. $\sigma_{int}^A$ is the temperature independent intrinsic anomalous Hall conductivity given by the temperature independent term c. The solid black line uses the parameters a$^\prime$, c$^\prime$ obtained from the linear fit of the $\rho^A_{yx}$ vs $\rho^2_{xx}$, for data presented in Fig. S5(a).}   \label{F3}
\end{figure*}

We measured the Hall resistivity of Tb166 as a function of magnetic field [$\rho_{yx}(B)$] between 1.8 and 300 K as shown by some representative data in
 Figure \ref{F3}(a). The zero field value of  $\rho_{yx}$ gives the anomalous Hall resistivity $\rho^A_{yx}$. At higher temperatures, $\rho_{yx} (B)$ follows the
 magnetization $M(B)$ and shows saturation behavior when the latter saturates, indicating that magnons play a decisive role in high-temperature 
transport. Below  60 K, although $M(B)$ shows a pronounced hysteresis,  the hysteresis in $\rho_{yx} (B)$ becomes unresolvable within the error of the Hall resistivity measurement and makes it difficult to extract  $\rho^A_{yx}$ unambiguously (see Fig. S4). In Fig. \ref{F3}(b) we show the anomalous Hall conductivity 
$\sigma^A_{xy} $= $\rho^A_{yx}/\rho_{xx}^2$ (which holds for $\rho_{yx}<<\rho_{xx}$ as is the case in TbMn$_6$Sn$_6$) and longitudinal conductivity $\sigma_{xx}$ = 1/$\rho_{xx}$ as a function of temperature. We can see clearly that  $\sigma^A_{xy} $ shows temperature 
dependence in the entire temperature range, complicating the extraction of the temperature independent intrinsic Hall conductivity. Not surprisingly, Fig. \ref{F3}(c) shows that $\sigma^A_{xy}$ is not linear in  $\sigma_{xx}^2$, as expected in Eq. \ref{Eq6b} [the black solid line in Fig. S5(a) is plotted using the parameters $a^\prime$, and $c^\prime$ obtained from a straight line fit to $\rho^A_{yx}$ vs $\rho_{xx}^2$].
Here, we want to point out that our $\rho_{yx}^A$ vs. $\rho_{xx}^2$ is similar to that reported in Ref. \cite{Yin2020} where Eq. \ref{6a} (equivalent of Eq. \ref{Eq6b}) is used to extract the intrinsic AHC (see Fig. S5). The poor quality of Eq. \ref{6a} fit, when plotted in the proper scale, was 
also noticed in Ref. \cite{Binghai}, who used instead Eq. \ref{Eq5a} with a non-zero 
$b$; this led to a considerably improved, but still far from perfect fit.

On the other hand, a closer look at the data in Fig. \ref{F3}(c) shows that
 $\sigma^A_{yx}$ can be extremely well described by 
the relation:
  \begin{equation}
   \sigma_{xy}^A = a\sigma_{xx}^2+d/\sigma_{xx}+c \label{Eq6}
   \end{equation}
 as shown by the red curve in Fig. \ref{F3}(c). Comparing this with Eq. \ref{Eq6b} we observe that the main difference with the conventional
expression is that the inverse term in $\sigma_{xx}$ (equivalently, the cubic term in Eq. \ref{Eq5}) cannot be neglected. We do $not$ observe this term to be significant in the sister compound Y166 \cite{Siegfried2022},
which strongly suggests that it is due to scattering off of Tb magnons, in agreement with the arguments above.

The $T$-dependent term in $\sigma_{yx}^A(T)$ is due to impurity/defect scattering (cf. Eq. \ref{G}); indeed, above 150 K, the data can also be fitted 
just to $\sigma_{xy}^A = c + d/\sigma_{xx}$ [green curve in Fig. \ref{F3}(c)] yielding about  the same value for $c=\sigma^A_{int},$
indicating that the inverse term in $\sigma_{xx}$ at higher temperatures is essential. The value of 
$\sigma^A_{int}$ obtained from the fitting is $140 \pm 1$ S cm$^{-1}$. Although extrinsic effects, primarily, side jumps can also
 contribute to $\sigma^A_{int}$, which complicates comparison with theoretical models, it is likely subdominant \cite{Yang2011}  and hence it is 
clear that TbMn$_6$Sn$_6$  has a significant intrinsic AHC. What makes Tb166 unique is the presence of the $\sigma^{-1}_{xx}$ term, indicating an unconventional transport scattering most likely coming from Tb magnons. We emphasize that the same physics that controls the spin-reorientation transition also defines this unusual AHC scaling. 

\begin{figure*}[!ht]
\begin{center}
\includegraphics[width=.8\linewidth]{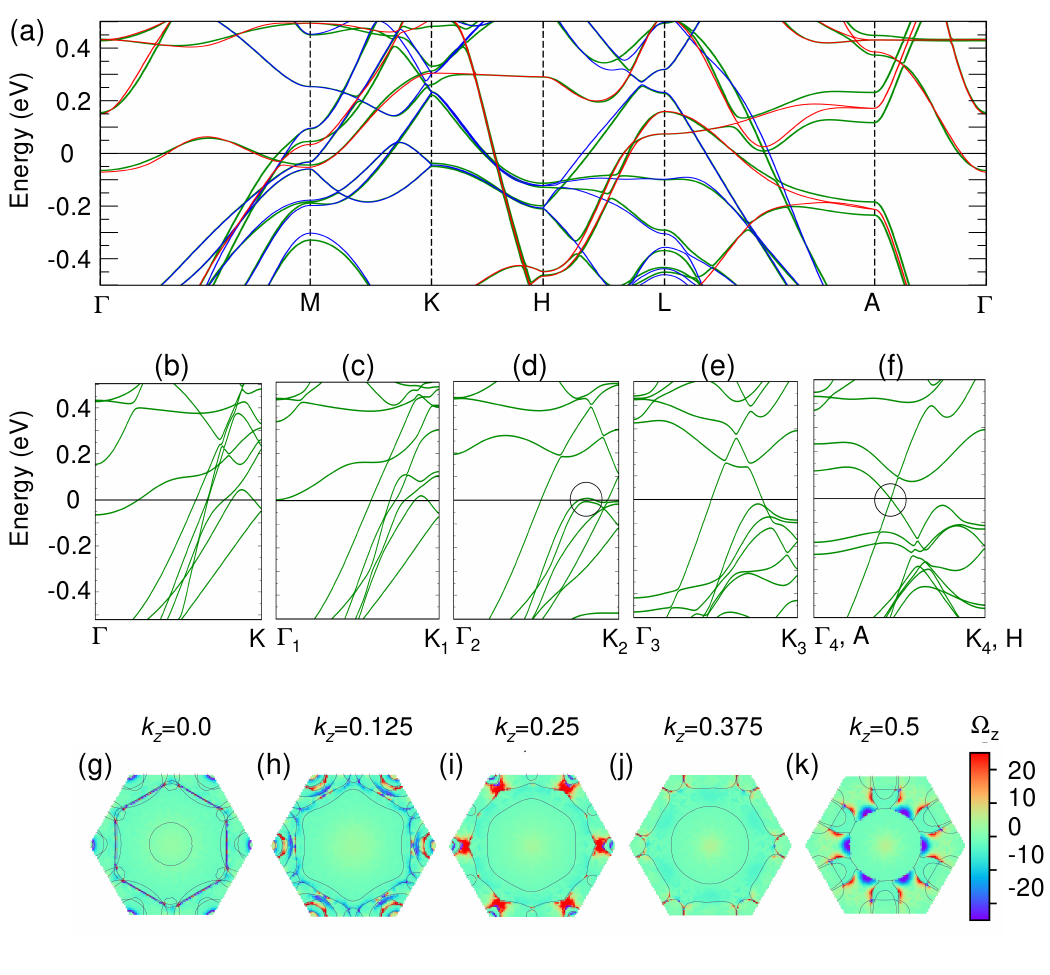}
\caption{The calculated band structure and band/momentum-resolved Berry curvature. (a) Spin-up (blue), spin-down (red) and spin-orbit (green) bands.
(b-f) Bands along the generalized $\Gamma$-K directions;  $\Gamma_i$-K$_i$ correspond to $k_z=0.125$, 0.25, and 0.375 r.l.u for $i=1-3$.
No-subscript points refer to actual  $\Gamma$-K ($k_z=0$), and $i=4$ to the AH direction  ($k_z=0.5$). (g-k) Berry curvature $\Omega_z$
at different points in the BZ. Panels from left to right correspond to the same $k_z$ and panels in (b-f). The largest positive contribution comes 
from $k_z\approx 0.25$, negative from $k_z\approx 0.5$; the corresponding parts of the band structure are encircled in (d), and (f), respectively.}\label{F4}
\end{center}
\end{figure*}

Next, we focus on the evaluation of intrinsic $\sigma^A_{int}$ and study it microscopically utilizing Density Functional Theory (DFT) and maximally localized Wannier functions (MLWF), as described in the Methods section.
Fig. \ref{F4}(a-f) shows the Wannier-interpolated electronic structure for TbMn$_6$Sn$_6$. We note that five Mn $d$ orbitals in the two spin directions per each 
of the two Kagome layers 
provide a multitude of Dirac lines along the  K-H, reminiscent of the single-orbital 2D tight-binding model. Many propitious features are washed out by interorbital hybridization, but 
 at least 8 are reasonably well expressed.  Of them, two spin-up DP are located at $-50$ and $+200$ meV below (above) the Fermi level.
In addition, there are a number of DPs not related to this TB model; for instance, two spin-up DPs occur at M, at $-50$ and $-200$ meV, plus, there
are about a dozen of accidental DPs all over the BZ, only a few of them relevant for AHC.

Two further DPs are noticeable: one is formed by the spin-down  $d_{z^2-1}$ ($a_{1g}$, in hexagonal notations) Mn orbital; at K it occurs at $\sim 0.7$ eV
above $E_F$, and extremely rapidly disperses down, crossing the Fermi level midway between K and H. The other one is derived from the $x^2-y^2\pm ixy$
orbital and is truly 2D. This is the orbital that was postulated in Ref. \cite{Yin2020} to be responsible for the large AHC. However, as was convincingly 
demonstrated in Ref. \cite{Ke}, in a charge-balanced system this band is located about 0.7 eV above the Fermi level, and cannot contribute to
AHC. This conclusion fully agrees with our independent calculations, as well as with recently published calculations from the Weizmann group~\cite{Binghai}. It is worth noting that this state, being very 2D, is unlikely to dominates the STM spectra, which are normally sensitive to states extending into the vacuum, and 
$k_z$-dispersive.

\begin{figure}[!ht]
\hspace*{-0.5cm}\includegraphics[width=1.1\columnwidth]{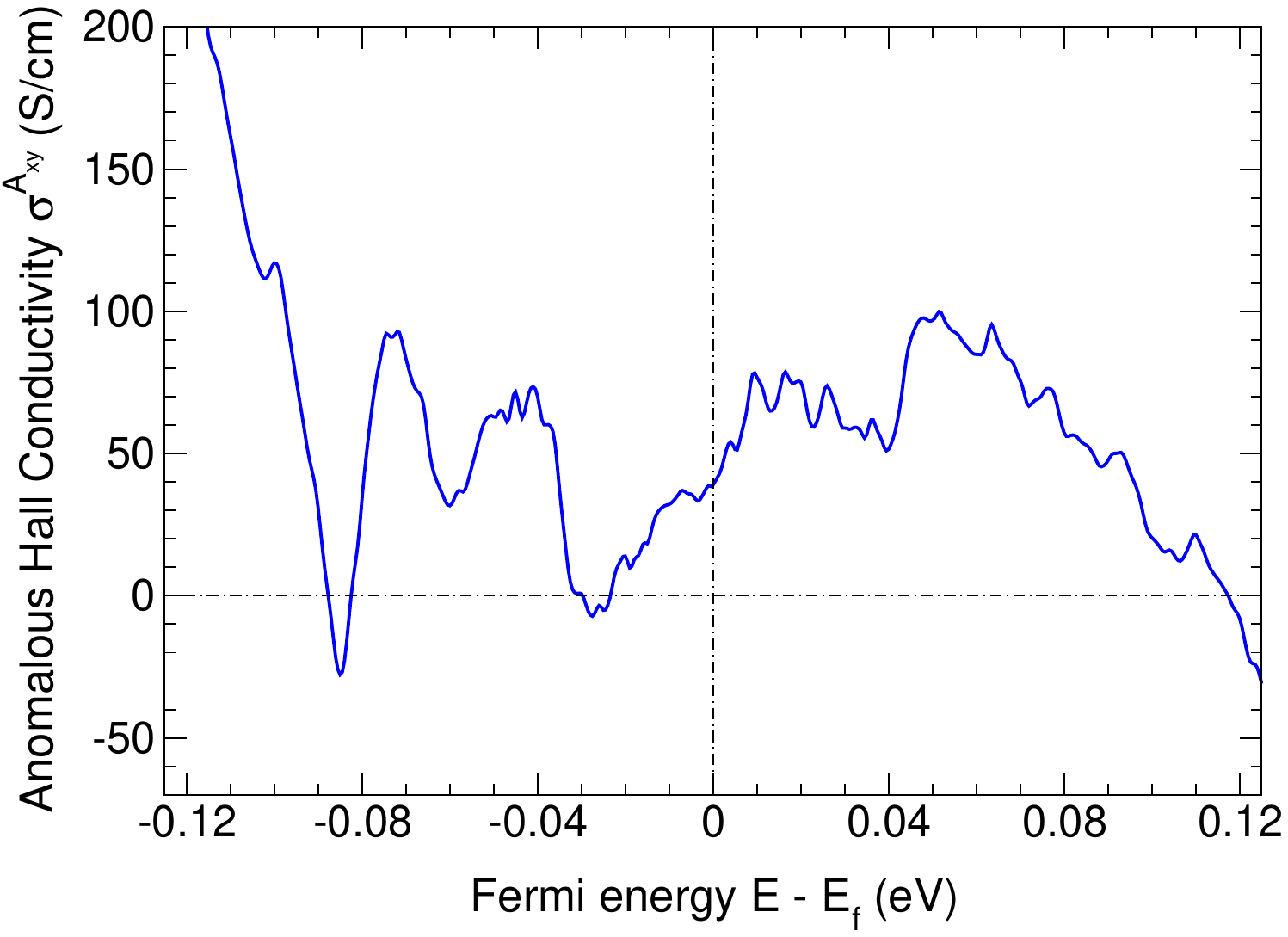}
\caption{The anomalous Hall conductivity $\sigma^{A}_{xy}$ as a function of $E-E_F$, obtained by applying iterative adaptive refinement, as described in the recent implementation for extraction of berry curvature.}\label{refined}
\end{figure}

In order to gain microscopic insight into the origin of the large AHC, we performed first principles calculations of the latter, 
using the code and methodology described in Ref. \cite{Vanderbilt1, Vanderbilt2}.
Our calculations have been extensively tested for convergence and precision with respect to k-mesh size and higher energy cutoffs.
The calculated  AHC ~\cite{Tsirkin2021} as a function of the position of the Fermi level is shown in Fig. \ref{refined}. At the theoretical $E_F$, it is 50 $\pm 1$ S cm$^{-1}$, in qualitative agreement
with the experiment, but underestimating the latter by some 60\%. For the Fermi level shifted up by $\sim 50$ meV, corresponding to the about 0.5
$e$/formula doping (note that we suggest that our samples are actually doped, it just illustrates the sensitivity of the final result to this parameter), the calculated $\sigma_{xy}$ is $\sim 80 $ S cm$^{-1}$, reducing the discrepancy to $\sim 40$\%. The remaining disagreement, apart from purely technical reasons, may 
be due to an additional extrinsic contribution, or, even more likely, to underestimation of the correlation effects on Mn. For instance, it was shown that in Sr$_2$RuO$_4$
the effective spin-orbit coupling in dynamical mean field calculations is enhanced by a factor of two compared to DFT \cite{AG}. It is worth noting that the AHC calculations are known to be extremely sensitive to the details of the computational setup, much more so than the band structure itself. In this regard, recent calculations of AHC \cite{Binghai}, while finding overall a rather similar structure of AHC, differ in some fine details near the Fermi energy. Having discussed this discrepancy with the authors of Ref. \cite{Binghai}, we have come together to the conclusion that this difference is due to slightly different computational setups. While this discrepancy does not affect our qualitative conclusion, it also serves as an independent test of the accuracy of the first principles calculations of AHC.

Valuable information can be derived from the color maps showing the contribution to the Berry curvature (essentially, to AHC) from different points in the 
BZ [Figs. \ref{F4}(g-k)]. 
Interestingly, the K-point, conjectured in Ref. \cite{Yin2020} to be the sole source of the AHC in Tb166, contributes very little, while large contributions come 
from two other  regions: one arises from complicated hybridization between the DP at K$_2$, emanating from the lowest-energy unoccupied DP at K,
 an avoided accidental tilted Dirac crossing between K$_2$ and $\Gamma_2$ (spin-up), and an accidental spin-down DP between A and L.

\section*{Conclusions}
We have presented combined experimental and theoretical studies 
of magnetic properties of TbMn$_6$Sn$_6$ in comparison to its f-electron-less analogue YMn$_6$Sn$_6$
 in order to clarify the following issues:
\begin{enumerate}
    \item Why is TbMn$_6$Sn$_6$  a collinear ferrimagnet while YMn$_6$Sn$_6$ is a spiral antiferromagnet?
    \item Why does TbMn$_6$Sn$_6$ experience a spontaneous magnetic reorientation transition $\approx 100$ K below it Curie temperature?
    \item What is the microscopic origin of the anomalous Hall effect in  TbMn$_6$Sn$_6$? 
    \item What are the scaling relations between the Hall conductivity and longitudinal resistivity in this compound?
\end{enumerate}

After a careful analysis of the experimental and computational data,
we have arrived at the following conclusions:
\begin{enumerate}
    \item Indirect exchange between the two Mn layers bridged by Tb, after integrating out Tb moments, leads to an effective ferromagnetic coupling, rendering an unfrustrated magnetic Hamiltonian. This is in contrast to the Y based analogues as well as other non-magnetic rare earths-based ones.
    \item The spin orientation is decided by the competition between the easy-axis Tb anisotropy and the easy plane Mn anisotropy. For fully ordered magnetic moments the former is much larger; however, the Tb 
    and Mn spins have very different thermal dynamics. The molecular (Weiss) field
    on the Mn site it much larger due to the very strong ferromagnetic in-plane Mn-Mn
    exchange, while the molecular field on Tb is much weaker. As a result, with the increase in temperature, Tb spins fluctuate more strongly as compared to those of Mn. Due to the fluctuations, Tb moment gets considerably reduced on approaching the Curie temperature ($T_C$); so does its magnetic anisotropy, which at $\sim 100$ K below $T_C$ drops below the Mn anisotropy. This scenario
     is qualitatively consistent with published neutron scattering and DFT results.
    \item The 3D nature of the Fermi surface is important in understanding the anomalous Hall effect in Tb166. The AHC is not related to a single K-point Dirac cone associated with the single band 2D tight-binding Kagome model, but rather comes from
    other bands.
    \item Magnon contribution, in addition to the impurity scattering and the temperature independent intrinsic AHC, may be important in the scaling relation between the AHC and the longitudinal conductivity in Tb166.
\end{enumerate}
\vskip 0.2 in
\section*{Acknowledgments}
The authors thank Predrag Nikolic, Patrick Bruno, Jairo Sinova, and especially Liqin Ke, and Binghai Yan, the authors of, respectively, Refs.~\cite{Ke,Binghai}, for insightful discussions. Crystal growth and characterization work at George Mason University is supported by the U.S. Department of Energy, Office of Science, Basic Energy Sciences, Materials Science and Engineering Division. H.B. acknowledges support from the NSF CAREER award DMR-2343536. I.I.M. acknowledges support from the U.S. Department of Energy through the Grant No. DE-SC0021089, and, at a later stage, from National Science Foundation through the Grant No. DMR-2403804. M.P.G. thanks the University Grants Commission, Nepal for the Collaborative Research Grants (Award No. CRG-78/79 S\&T-03), and acknowledges the Alexander von Humboldt Foundation, Germany for the equipment grants. M.P.G. thanks Manuel Richter for the fruitful discussions and Ulrike Nitzsche for the technical assistance. S.S.T. acknowledges funding by MCIN/AEI/10.13039/501100011033 through Grant No. PID2021-129035NB-I00, by European Union (Grants No. H2020-MSCA-COFUND-2020-101034228- WOLFRAM2, No. ERCStG-Neupert757867-PARATOP], and by Grant No. PP00P2-176877 from the Swiss National Science Foundation. X.L. acknowledges support from the China Scholarship Council (CSC).

\section*{References}


%

\widetext
\begin{center}
\pagebreak
\textbf{\large Supplementary Information}
\end{center}

\setcounter{equation}{0}
\setcounter{figure}{0}
\setcounter{table}{0}
\setcounter{page}{1}
\makeatletter
\renewcommand\thesection{S\arabic{section}}
\renewcommand{\theequation}{S\arabic{equation}}
\renewcommand{\thetable}{S\arabic{table}}
\renewcommand\thefigure{S\arabic{figure}}
\renewcommand{\theHtable}{S\thetable}
\renewcommand{\theHfigure}{S\thefigure}

\section*{Supplementary Note 1: Magnetic Anisotropy Energy}
We derive our expression for the magnetic anisotropy energy for the collinear arrangement of the moments on the Tb- and Mn-sublattices on the classical (Langevin) level. We first write the standard partition function for the external magnetic field $H$ parallel to the $z$-axis assuming the anisotropy is a small correction to the Zeeman term:

\begin{align}
\mathcal{Z} & =e^{-\beta F}=\int_0^{2\pi}d\varphi\int_0^\pi e^{-(-HM_z)\beta}sin\theta~d\theta \label{SA}\\
& = 2\pi\int_0^\pi e^{\beta HMcos\theta}sin\theta~d\theta,  \label{SB}\\
& = \frac{4 \pi~sinh(\beta HM)}{\beta HM}, \label{SC}
\end{align}

where $\beta = k_{B}T$ with $k_{B}$ being the Boltzmann constant, $M$ is the total magnetization, and $\theta$ is the angle between $M$ and $H$.
\\
The expectation value for the total magnetic moment is:

\begin{align}
   m &=\frac{1}{\mathcal{Z}}\int_0^{2 \pi} d \varphi \int_0^\pi M~cos\theta~e^{\beta HMcos\theta} sin\theta~d\theta \\
    &=Mcoth(\beta HM)-\frac{1}{\beta H}
\end{align}

By defining the hexagonal axis $c$ to lie along field direction $z$, then the magnetic anisotropy energy can written as
\begin{equation}
\mathcal{H}_{anis} = K M_z^2 = K M^2 \cos^2{\theta}
\end{equation}

where $K$ is the $2^{nd}$ order anisotropy coefficient.\bigskip 

For the easy-axis, parallel to $H$, the expectation value is

\begin{align}
    E_\| & = \frac{1}{\mathcal{Z}}\int_0^{2\pi} d\varphi \int_0^\pi K M^2~cos^2\theta~e^{\beta HMcos\theta} sin\theta~d\theta \\
       &=KM^2-\frac{2K}{\beta H}m
\end{align}

For the easy axis perpendicular $H$, the expectation value is

\begin{align}
    E_\bot & = \frac{1}{\mathcal{Z}}\int_0^{2\pi} cos^{2}\varphi d\varphi \int_0^\pi K M^2~sin^2\theta~e^{\beta HMcos\theta} sin\theta~d\theta \\
    &=\frac{K}{ \beta H}m\\
\end{align}

The effective magnetic anisotropy energy is thus

\begin{align}
   MAE = E_\|-E_\bot = KM^2 -  \frac{3K}{\beta H}m
\end{align}

To eliminate $H$, we solve the transcendental equation $m$ = $Mcoth(\beta HM)$-$1/\beta H$ for $H$, by rewriting it as

\begin{align}
\mu = coth(h)-\frac{1}{h},
\end{align}

where $\mu$ =$m/M$ and $h$ = $\beta HM$.\bigskip 

One finds solutions for $\mu \to 0$ and $\mu \to 1$ by interpolating between the two:
\begin{equation}
h = \mu (2 - \mu + \frac{1}{1-\mu}) \implies H = \frac{m}{\beta M^2} (2 - \frac{m}{M} + \frac{M}{M-m})
\end{equation}
\begin{equation}
MAE = KM^2 - 3K \frac{M^2}{2 - \frac{m}{M} + \frac{M}{M-m}} = Km^2 \frac{M^2}{3M^2 - 3Mm + m^2}
\end{equation}

Note that $MAE \to Km^2$ at $T = 0$, where $m = M$, and $MAE \to \frac{1}{3}KM^2$ near $T_C$, where $M \to 0$.

\begin{figure}[th]
\begin{center}
\includegraphics[scale=.6]{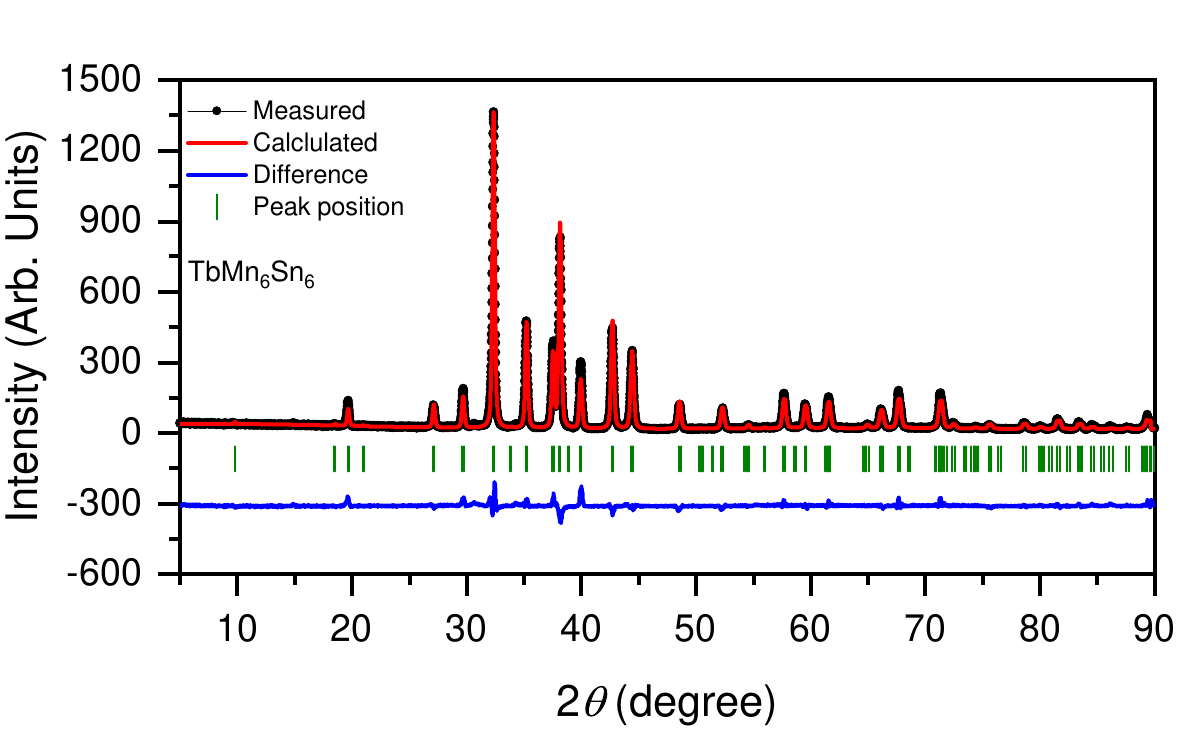}
\end{center}
\caption{Rietveld refinement of the X-ray powder pattern of TbMn$_6$Sn$_6$ measured at room temperature.} \label{FigXRD}%
\end{figure}
\begin{table}[!ht]
\caption{Selected data from Rietveld refinement of powder X-ray diffraction collected on ground crystals of TbMn$_6$Sn$_6$. Atomic coordinates are 0, 0, 0 for Tb; 0, $\frac{1}{2}$, $z$ for Mn; 0, 0, $z$ for Sn(1); $\frac{1}{3}$, $\frac{2}{3}$, $\frac{1}{2}$ for Sn(2); and $\frac{1}{3}$, $\frac{2}{3}$, 0 for Sn(3).}\label{T0}
\begin{center}
\par%
\begin{tabular}
[c]{ll}\hline
   Space group                           &      \textit{P6/mmm } (No. 191)             \\
   Unit cell parameters                &     \textit{a} = 5.5384(3) \AA               \\
                                                    &     \textit{c} =  9.0325(6)  \AA          \\
 \textit{R}$_{WP}$                     &     14.6 \%                               \\
  \textit{R}$_{B}$                     &     8.61 \%                               \\
   \textit{R}$_{F}$                           &     7.43  \%                               \\
  Mn $z$ coordinate                  &     0.24803(31)                           \\
  Sn(1) $z$ coordinate              &     0.33325(26)                    \\
 \hline
\end{tabular}
\end{center}
\end{table}

\begin{figure}[th]
\begin{center}
\includegraphics[scale=.35]{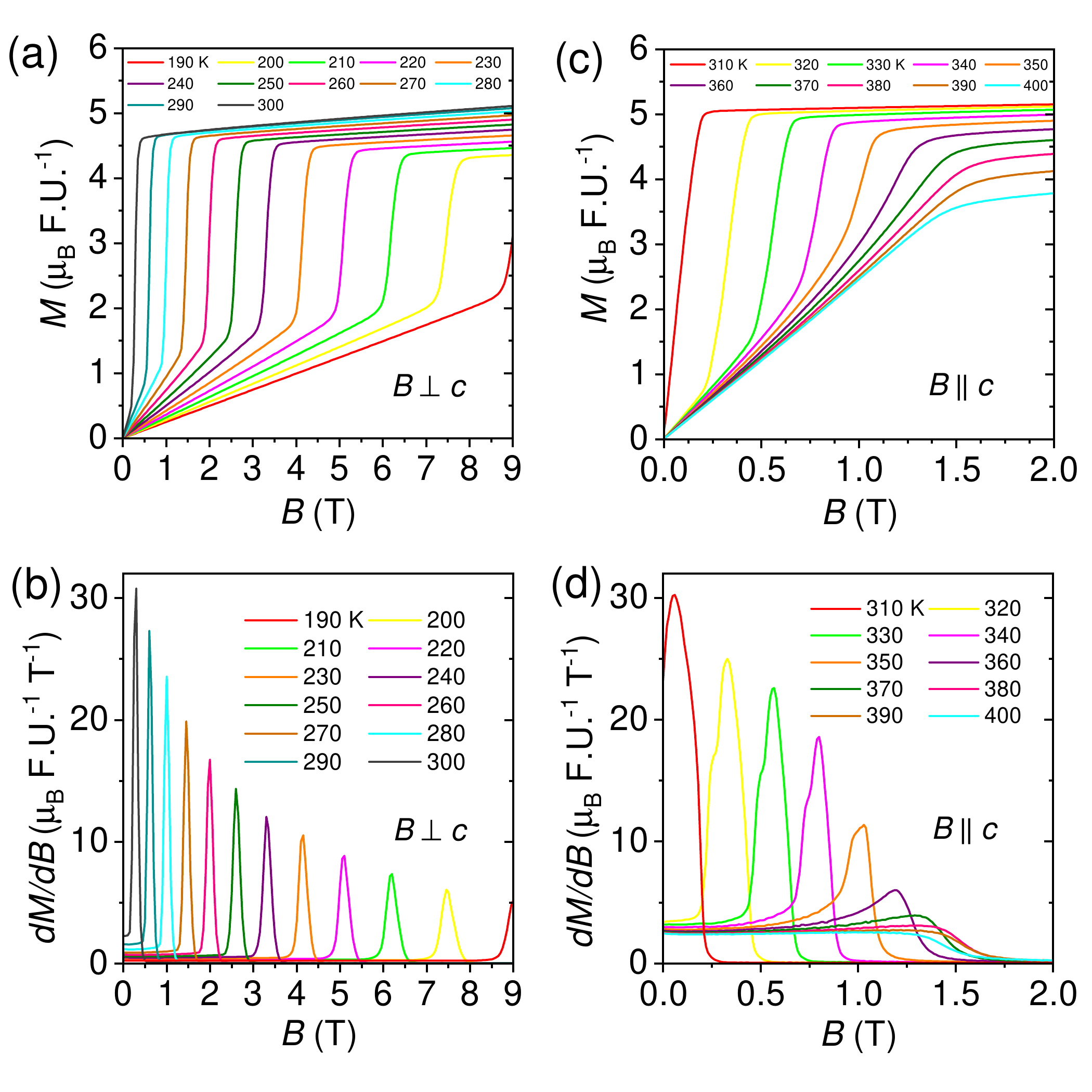}
\end{center}
\caption{ First Order Magnetization Process (FOMP) in TbMn$_6$Sn$_6$. (a) Magnetic field dependence of magnetization $M(B)$ with $B\perp c$ below the spin-reorientation transition temperature at zero-field $T_{sr}$. (b) Magnetic field dependence of derivative of magnetization $dM/dB(B)$ with $B\perp c$ below $T_{sr}$. (c) $M(B)$ with $B\|c$ above $T_{sr}$.  (d) $dM/dB(B)$ with $B\|c$ above $T_{sr}$.} \label{SMH2}%
\end{figure}

\begin{figure}[th]
\begin{center}
\includegraphics[scale=.4]{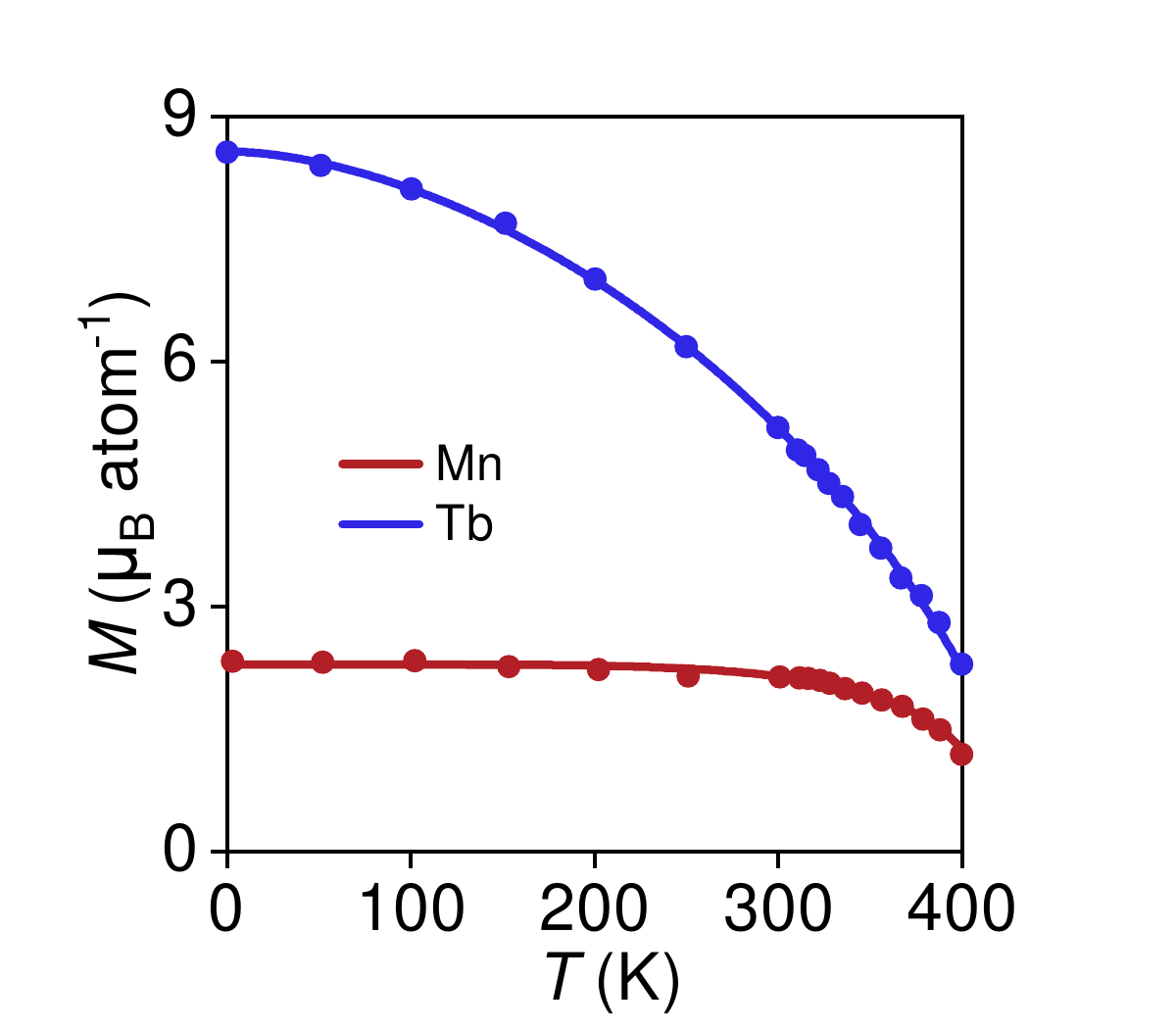}
\end{center}
\caption{ Fit of Brillouin curves (solid lines) to experimental data (points) for the temperature dependence on the magnetic moments on Mn and Tb obtained from a previous neutron powder diffraction study [25].} \label{SMH3}%
\end{figure}

\begin{figure}[th]
\begin{center}
\includegraphics[scale=1]{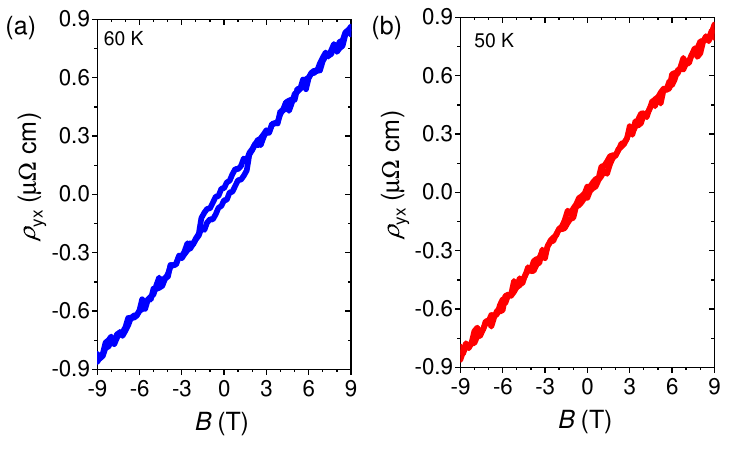}
\end{center}
\caption{Hall resistivity as a function of magnetic field at a) 60 K and , b) 50 K. Note that there is a clear hysteresis in the Hall resistivity at 60 K, which allows to extract the anomalous Hall resistivity ($\rho_{yx}^A$) unambiguously. The hysteresis disappears (or is within the error bar of the resistivity measurement) at 50 K which does makes it difficult to get $\rho_{yx}^A$ from these data at and below 50 K.} \label{SMH4}%
\end{figure}

\begin{figure}[th]
\begin{center}
\includegraphics[scale=.32]{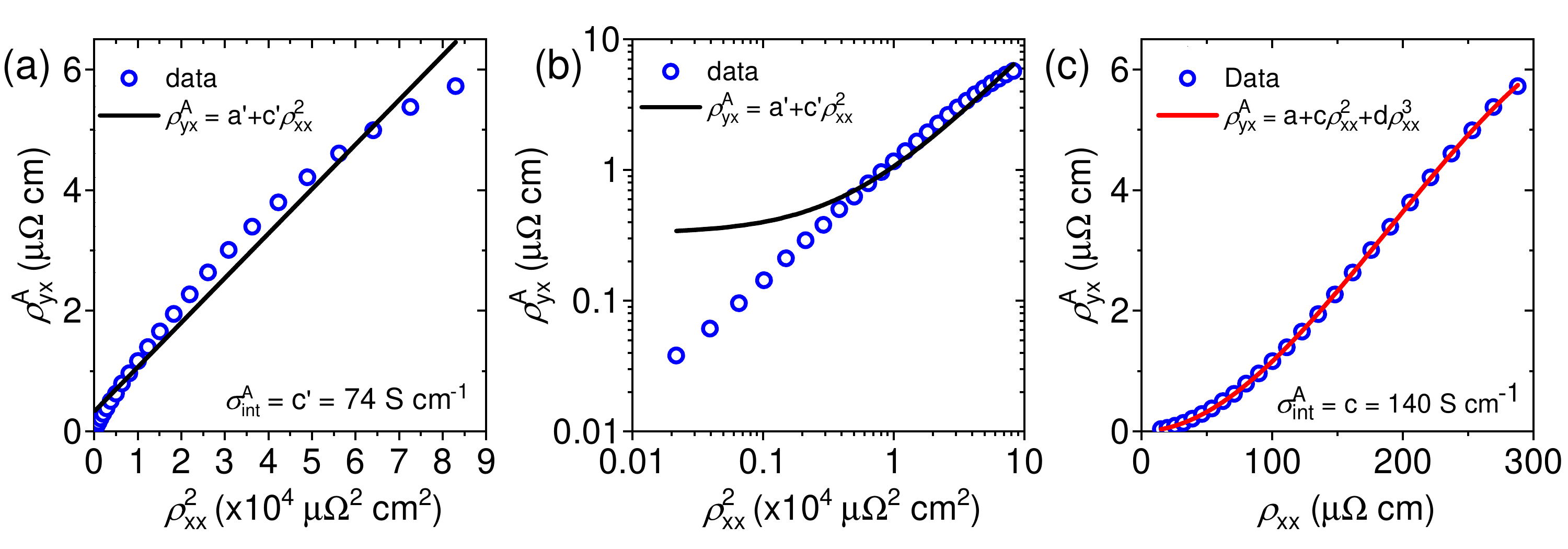}
\end{center}
\caption{a) Anomalous Hall resistivity ($\rho_{yx}^A$) as a function of square of longitudinal resistivity ($\rho_{xx}^2$). Black line is a straight line fit to the data which shows that $\rho_{yx}^A$ has a systematic deviation from the $\rho_{xx}^2$ dependence, which is more clearly seen in the log-log plot presented in Panel (b). Here, c$^\prime$, the coefficient of $\rho_{xx}^2$, gives the intrinsic anomalous Hall conductivity ($\rho_{xy}^A$), which is found to be 74 S cm$^{-1}$ from this fit.  c) $\rho_{yx}^A$ vs. $\rho_{xx}$. The red line shows a fit to a+c$\rho_{xx}^2$+d$\rho_{xx}^3$, which shows that the cubic term is necessary to describe the  $\rho_{xx}$ dependence of $\rho_{yx}^A$. $\rho_{xy}^A$ obtained from this fitting is 140 S cm$^{-1}$.} \label{SMH5}%
\end{figure}
\clearpage

\end{document}